\documentclass[pre,twocolumn]{revtex4}
\usepackage{epsfig}
\usepackage{bm}
\usepackage{amsmath}
\usepackage{hyperref}
\usepackage{breakurl}

\begin{document}

\title{Strong chaotic fluctuations of laser beams propagating through temperature inhomogeneities in the atmosphere}

\author{A. Bershadskii}

\affiliation{
ICAR, P.O. Box 31155, Jerusalem 91000, Israel
}

\begin{abstract}

The atmospheric temperature inhomogeneities are the main cause of the index of refraction variations resulting in strong fluctuations of intensity of the laser beams propagating through the atmosphere. It is shown that the Kolmogorov-Bolgiano-Obukhov phenomenology of the thermal (buoyancy driven) convection can effectively describe this process in the frames of the distributed chaos approach. Results of the direct numerical simulations and measurements in the urban atmosphere (based on the laser beams with the wavelengths of 532nm and 690nm) have been used in order to support the theoretical considerations.

\end{abstract}

\maketitle

\section{Inroduction}

 It is well known that in the visible and infra-red ranges the atmospheric temperature inhomogeneities (fluctuations) are the main cause for fluctuations of the intensity of the laser beams propagating through the atmosphere (the absorption and scattering are as a rule the second order effects) \cite{andr}. This phenomenon is related to the fluctuations of the index of refraction due to the temperature inhomogeneities (optical turbulence) and can have far reaching consequences on all applications requiring transmission of laser beam through the atmosphere: laser radar, remote sensing, optical communications and imaging etc. (see, for instance, Ref. \cite{bos} and references therein).\\
 
 One of the main sources of the temperature fluctuations in the atmosphere is the thermal (buoyancy driven) convection. The temporal and spatial scales of the atmospheric convection are usually corresponding to the turbulent fluid dynamics that makes description of the processes very complex \cite{wyn}. The distributed chaos approach (a generalization of the deterministic chaos approach) allows significant simplification of the description and its combination with the Kolmogorov-Bolgiano-Obukhov phenomenology \cite{my} provides a straightforward method for explanation and interpretation of the data obtained in the direct numerical simulations and atmospheric measurements. It should be also noted that the chaotic nature of the strong fluctuations of the laser beam intensity in the atmosphere is a very significant stabilising factor for the chaos based self-synchronizing communication systems \cite{rvi}. \\
 
 Section II of present paper provides a short relevant for our purpose description of the thermal (buoyancy driven) convection in the Boussinesq approximation and the Kolmogorov-Bolgiano-Obukhov phenomenology. Section III provides description of the distributed chaos approach based on this phenomenology and comparison with results of direct numerical simulations. In section IV the above described theoretical consideration has been compared with the results of measurements of the laser beams propagation through the urban atmosphere.

\section{Thermal convection }

  The thermal (buoyancy driven) convection in the Boussinesq approximation is described by the system of equations \cite{kcv}
$$
\frac{\partial {\bf u}}{\partial t} + ({\bf u} \cdot \nabla) {\bf u}  =  -\frac{\nabla p}{\rho_0} + \sigma g \theta {\bf e}_z + \nu \nabla^2 {\bf u}   \eqno{(1)}
$$
$$
\frac{\partial \theta}{\partial t} + ({\bf u} \cdot \nabla) \theta  =  S  \frac{\Delta}{H}e_z u_z + \kappa \nabla^2 \theta, \eqno{(2)}
$$
$$
\nabla \cdot \bf u =  0 \eqno{(3)}
$$
where ${\bf u}$ is the velocity, $\theta$ is the temperature fluctuations, $p$ is the pressure, ${\bf e}_z$ is the a unit vector along the gravity, $g$ is the gravity acceleration, $H$ is the distance and $\Delta$ is the temperature difference between the layers, $\nu$ is the viscosity and $\kappa$ is the thermal diffusivity, $\rho_0$ is the mean density, $\sigma$ is the thermal expansion coefficient.  For the unstable (Rayleigh-B\'{e}nard) stratification the coefficient $S=+1$ and for the stable stratification the coefficient $S=-1$.\\   

  In the non-viscous and non-diffusive case ($\nu=\kappa=0$ ) this system has an invariant \cite{kcv}
$$
\mathcal{E} = \int_V ({\bf u}^2 -S\sigma g \frac{H}{\Delta}\theta^2) ~ d{\bf r},   \eqno{(4)}
$$    
where $V$ is the volume of the domain under consideration.
  
  A generalization of the Kolmogorov-Bolgiano-Obukhov approach \cite{my} for the inertial-buoyancy range can relate the characteristic temperature fluctuations $\theta_c$ and the characteristic frequency $f_c$
$$
\theta_c\propto  (\sigma g)^{-1} \varepsilon^{1/2} f_c^{1/2}, \eqno{(5)}
$$
where 
$$
\varepsilon = \left|\frac{d\langle{\bf u}^2 -S\sigma g \frac{d}{\Delta}\theta^2 \rangle}{dt}\right| \eqno{(6)}
$$
denotes the (generalized) dissipation rate and the $\langle ... \rangle$ denotes averaging over the spatial domain.\\

\section{Distributed chaos} 

   The exponential frequency spectrum 
$$ 
E(f) \propto \exp(-f/f_c)  \eqno{(7)}
$$
 is a typical spectrum for the deterministic (bounded and smooth) dynamical systems (see, for instance, Refs. \cite{fm},\cite{oh} and references therein). 
 
 A simplified model of the Rayleigh-B\'{e}nard convection (Lorenz system \cite{lorenz})
$$
\frac{dx}{dt} = \sigma (y - x),~~      
\frac{dy}{dt} = r x - y - x z, ~~
\frac{dz}{dt} = x y - b z,                
$$
for instance, exhibits such spectrum for the parameters $b = 8/3,~ r = 28.0, ~\sigma=10.0$. Figure 1 shows the power spectrum of $z$-component. The dashed straight line corresponds to the Eq. (7) in the semi-logarithmic scales. The exponential power spectrum is also observed for some real atmospheric processes with the chaotic behaviour (see, for instance, Ref. \cite{b1}). \\

  For more complex situations the parameter $f_c$ can fluctuate and it is necessary to consider an ensemble average over this parameter to compute the power spectrum 
$$
E(f) = \int P(f_c) ~\exp-(f/f_c)~ df_c, \eqno{(8)}
$$  

  In the frames of the Kolmogorov-Bolgiano-Obukhov phenomenology the distribution $P(f_c)$ can be readily found from the Eq. (5) in the case of  Gaussian distribution of the characteristic temperature fluctuations $\theta_c$ 
$$
P(f_c) \propto f_c^{-1/2} \exp-(f_c/4f_{\beta})  \eqno{(9)}
$$
where $f_{\beta}$ is a constant parameter.
\begin{figure} \vspace{-1.4cm}\centering
\epsfig{width=.45\textwidth,file=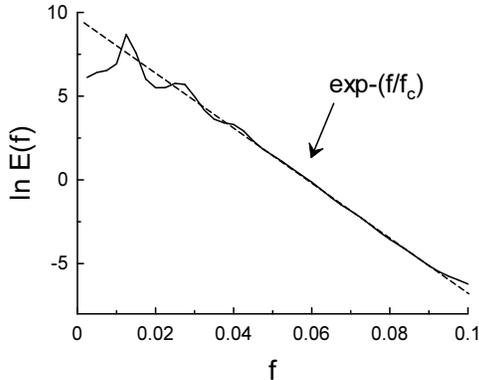} \vspace{-4.2cm}
\caption{Power spectrum of $z$-component for the Lorenz model.} 
\end{figure}
\begin{figure} \vspace{-2.75cm}\hspace{-1cm}\centering
\epsfig{width=.52\textwidth,file=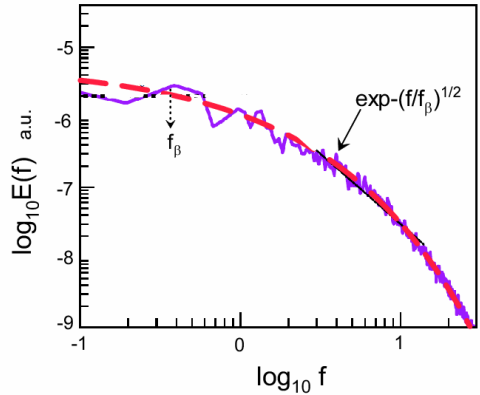} \vspace{-4.7cm}
\caption{Power spectrum of the temporal temperature fluctuations for the DNS of the the Rayleigh-B\'{e}nard convection. } 
\end{figure}

     Substituting the Eq. (9) into the Eq. (8) we obtain
$$
E(f) \propto \exp-(f/f_{\beta})^{1/2}  \eqno{(10)}
$$

 Figure 2 shows a frequency power spectrum for the temperature fluctuations obtained in direct numerical simulations (DNS) of the system Eqs. (1-3) (the Rayleigh-B\'{e}nard convection conditions). The spectral data have been shown in the log-log scales and were taken from Fig. 7 of the the Ref. \cite{kv}. The measurements in these DNS were made using real-space probes located at the centre of a cubical box. The mean velocity at the centre of the cube has zero value. Therefore, the measurements with the real-space probes give real temporal (frequency) spectrum \cite{kv}. The Prandtl number was taken $Pr = 1$ (i.e. of the order usual for the air at the normal conditions) and the Rayleigh number was taken $Ra = 10^8$. 

   The dashed curve in the Fig. 2 indicates the stretched exponential spectrum Eq. (10) and the dotted vertical arrow indicates location of  the frequency $f_{\beta}$. One can see that the distributed chaos is tuned to the coherent low-frequency oscillations with the frequency $f_{\beta}$.  \\
 
\section{Propagation of the laser beams in the atmosphere}

  The most significant effects on the propagating laser beams in the atmosphere are usually those caused by the atmospheric temperature fluctuations, resulting in the index of refraction variations (see, for instance, Refs. \cite{andr},\cite{andrew} and references therein). If the probability distribution $P(f_c)$ for the intensity fluctuations of the laser beams has the same functional form as for the temperature fluctuations (at least for the moderate to strong intensity), then the spectra of the intensity fluctuations of the laser beams can be directly used for the estimations of the temporal (frequency) spectra of the temperature fluctuations spatially integrated along the optical path of the laser beam propagating through the atmosphere.\\
  
  Propagation of the laser beams in the atmosphere over urban terrain is the most interesting phenomenon in the present context.
\begin{figure} \vspace{-2.1cm}\centering
\epsfig{width=.45\textwidth,file=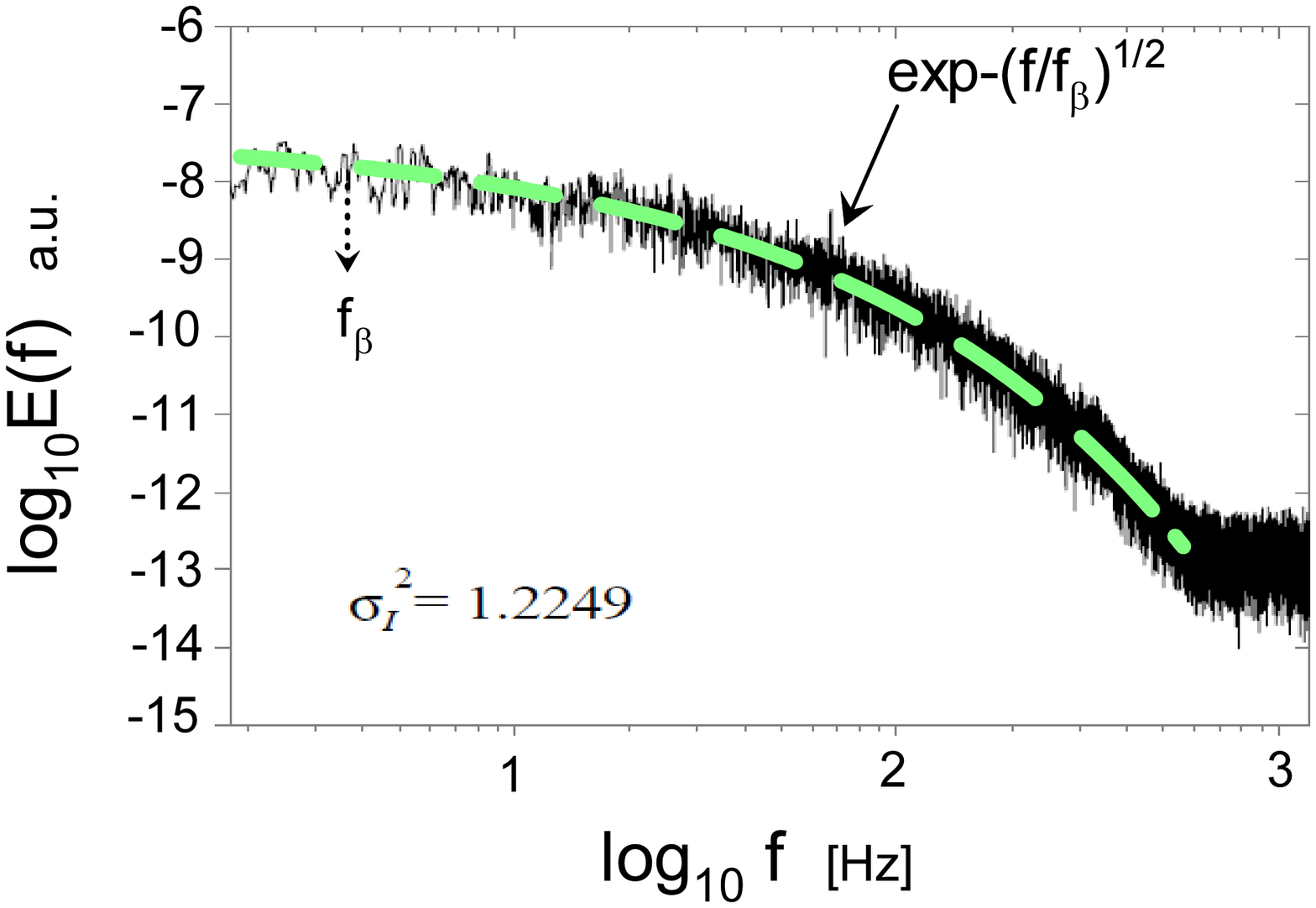} \vspace{-4.1cm}
\caption{Power spectrum of the intensity fluctuations of a laser beam at a receiver.} 
\end{figure}
\begin{figure} \vspace{-0.5cm}\centering
\epsfig{width=.48\textwidth,file=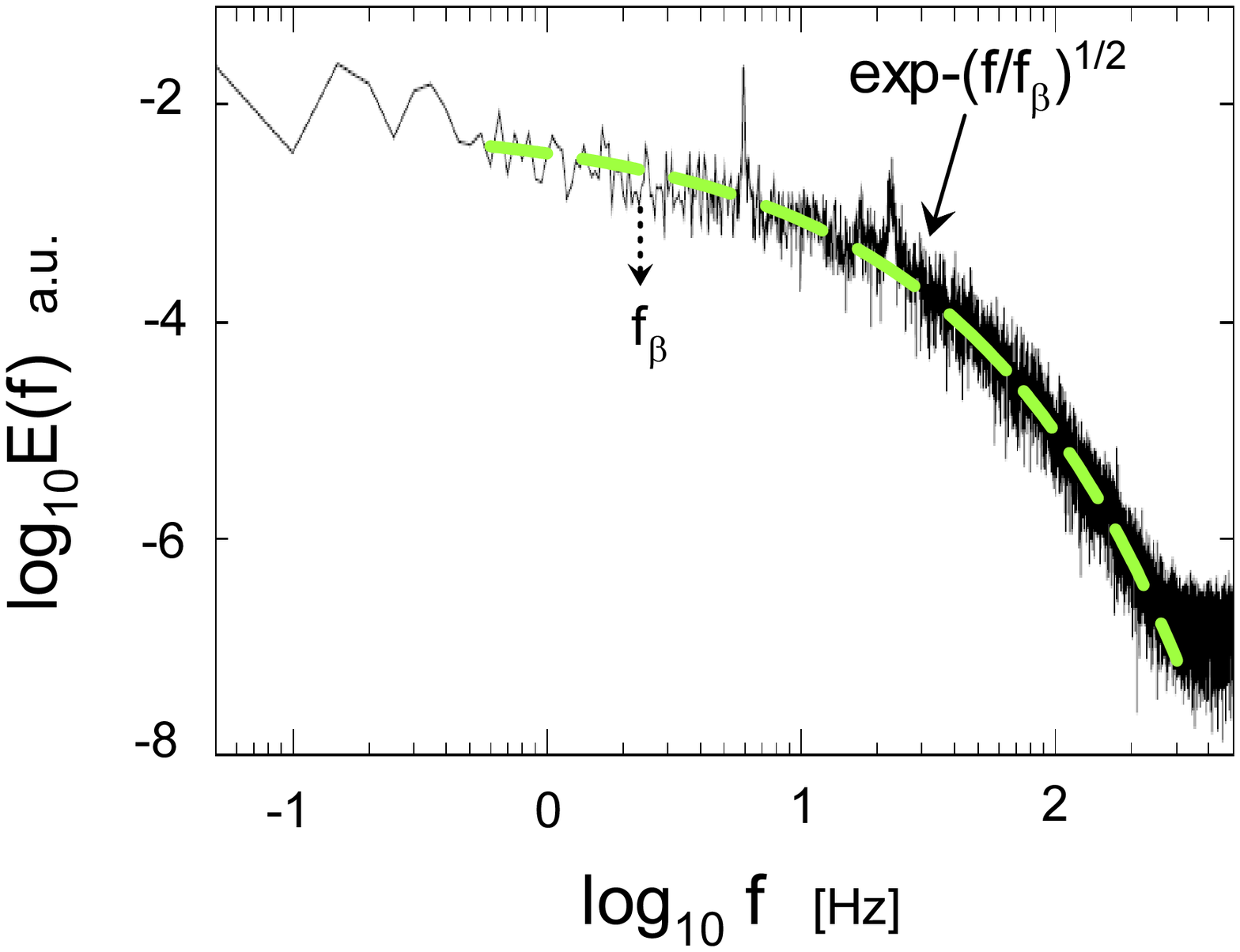} \vspace{-5.1cm}
\caption{Power spectrum of the intensity fluctuations of a laser beam at a receiver.} 
\end{figure}
  Figure 3 shows power spectrum of the intensity fluctuations of a laser beam (propagating through the urban atmosphere) at a receiver. The optical path through the urban atmosphere (the altitude from 30 to 55 m) had length $\simeq 11.8$ km. A laser with the wavelength of 532nm has been used in this experiment. The spectral data were taken from Fig. 9d of the Ref. \cite{jia}. The dashed curve indicates correspondence to the stretched exponential spectrum Eq. (10). The dotted arrow indicates position of the $f_{\beta}$.\\

 Figure 4 shows power spectrum of the intensity fluctuations of a laser beam at a receiver in another similar experiment \cite{rvi}. The optical path through the urban atmosphere (the altitude from 38.1 to 65.5 m) had length $\simeq 5$ km. A laser with the wavelength of 690nm has been used in this experiment. The spectral data were taken from Fig. 2b of the Ref. \cite{rvi}. The dashed curve indicates correspondence to the stretched exponential spectrum Eq. (10). The dotted arrow indicates position of the $f_{\beta}$.

  The experiment described in the Ref. \cite{rvi} was performed in order to study influence of the atmosphere caused errors on stability of the laser communication link based on a self-synchronizing chaos. The carrier of the transmitted binary information in this case is a chaotic sequence of short-term impulses. Because of the deterministic nature of the transmitted chaotic signal the two coupled systems (transmitter and receiver) can self-synchronize, that allows good reproducing of the transmitted information \cite{fy},\cite{pc}. The spectrum shown in the Fig. 4 corresponds to a non-modulated signal. It is reported in the Ref. \cite{rvi} that despite the strong intensity fluctuations, caused by the atmospheric temperature inhomogeneities, the laser communication method supports stable communication. The chaotic nature of the strong intensity fluctuations of the laser beams, caused by the atmospheric temperature inhomogeneities, can be a reason for this stability. 
  
  On the other hand, due to the direct relationship between the atmospheric temperature fluctuations and the fluctuations of the laser beam intensity the above described atmospheric measurements can be used for investigation of the pure temporal behaviour of the atmospheric thermal (buoyancy driven) convection, that can be hardly done using conventional methods (see, for instance, the Ref. \cite{kv} and references therein).

\end{document}